\documentclass[12pt]{article}

\usepackage{url}
\usepackage{lipsum}
\usepackage{algorithm}
\usepackage{algpseudocode}
\usepackage{xr-hyper}
\usepackage{amsthm}
\usepackage{amsfonts}
\usepackage{fullpage}
\usepackage{footnote}
\usepackage{tabu}
\usepackage{booktabs}
\usepackage{scribe}
\usepackage{listings}
\usepackage[mathscr]{euscript}

\usepackage{natbib}

\usepackage{stmaryrd}
 
\usepackage{graphicx}
\usepackage{subfigure}
\usepackage{multirow}
\usepackage{amsmath}
\usepackage{xcolor}
\usepackage[margin=0.75in]{geometry}

\theoremstyle{thmstyleone}%
\theoremstyle{thmstyletwo}%
\theoremstyle{thmstylethree}%

\begin{document}




 \title{\bf BAMITA: Bayesian Multiple Imputation for Tensor Arrays}
  \author{Ziren Jiang$^1$, Gen Li$^{2}$, and Eric F. Lock$^{1}$\hspace{.2cm}\\ \\
   $^1$Division of Biostatistics and Health Data Science, University of Minnesota \\
    $^2$Department of Biostatistics, University of Michigan}
    \date{}
  \maketitle





\abstract{Data increasingly take the form of a multi-way array, or tensor, in several biomedical domains. Such tensors are often incompletely observed. For example, we are motivated by longitudinal microbiome studies in which several timepoints are missing for several subjects.  There is a growing literature on missing data imputation for tensors. However, existing methods give a point estimate for missing values without capturing uncertainty.  We propose a multiple imputation approach for tensors in a flexible Bayesian framework, that yields realistic simulated values for missing entries and can propagate uncertainty through subsequent analyses. Our model uses efficient and widely applicable conjugate priors for a CANDECOMP/PARAFAC (CP) factorization, with a separable residual covariance structure.  This approach is shown to perform well with respect to both imputation accuracy and uncertainty calibration, for scenarios in which either single entries or entire fibers of the tensor are missing. For two microbiome applications, it is shown to accurately capture uncertainty in the full microbiome profile at missing timepoints and used to infer trends in species diversity for the population.  Documented R code to perform our multiple imputation approach is available at \url{https://github.com/lockEF/MultiwayImputation}.  }

\maketitle

\def\spacingset#1{\renewcommand{\baselinestretch}%
{#1}\small\normalsize} 
\spacingset{1.25}
\section{Introduction}
Multiway data, also known as a multidimensional array or a tensor, has become more prevalent in biomedical signal processing and other diverse fields. Tensors can effectively represent real-world datasets that have multiple dimensions or aspects. 
However, a common challenge associated with multi-way data is missing data, which can occur either on an entry-wise basis (i.e., certain entries have missing data) or a fiber-wise basis (i.e., entire modes have missing data). 
For example, in this manuscript we consider microbiome studies in which the abundance of microbial taxa are collected longitudinally over several timepoints,  for several subjects.  The resulting data can be represented as a 3-way tensor: subjects $\times$ taxa $\times$ time points; however, the microbiome profiles (abundance across taxa) are entirely missing at several time points for several of the subjects in both studies, resulting in fiber-wise missingness.  Our goal is to enable analysis of the full tensor by imputing missing data, while also accurately capturing uncertainty in the imputed fibers. Accounting for this uncertainty is particularly critical for the validity of subsequent  inferences, such as for population trends in species diversity over time.   

The Tucker decomposition \cite{tucker1966some} and CANDECOMP/PARAFAC (CP) decomposition \citep{carroll1970analysis} are two classic formulations to decompose tensor data, and can be viewed as extensions of the singular value decomposition for a matrix.  The Tucker decomposition decomposes a given tensor into a core tensor multiplied by matrices corresponding to each mode. The CP decomposition, which can be viewed as a constrained case of the Tucker decomposition, decomposing the tensor into a sum of rank-1 tensors.  \citet{kolda2009tensor} gives a very comprehensive review of tensor decompositions and their application in various contexts.

Tensor decomposition methods are useful to uncover the underlying structure in a tensor, and have been used to impute missing elements. \citet{acar2011scalable} describe imputation approaches based on the CP decomposition, and   \citet{chen2013simultaneous} describe imputation approaches based on a Tucker decomposition. Several related extensions incorporate penalization or smoothing into estimation of the tensor decomposition for missing data imputation \citep{yokota2016smooth,tan2013tensor, wu2018fused}.  An alternative approach, first proposed by \citet{liu2012tensor}, builds upon ideas in low-rank matrix completion \citep{mazumder2010spectral} by minimizing the tensor trace norm to transform the problem into a convex optimization problem.

The aforementioned imputation approaches are all deterministic, in that they only produce a single-point estimate with no uncertainty for the imputed values. 
A Bayesian framework is attractive in this context, because it can accommodate the collective uncertainty of the underlying factors that are combined in a low-rank decomposition.  There is an extensive literature on low-rank Bayesian modeling for tensors, particularly in the regression context \citep{hoff2015multilinear,guhaniyogi2017bayesian,wang2024bayesian}. However there is little work on Bayesian approaches to multiple imputation for tensors. \citet{chen2019bayesian} proposed a Bayesian tensor decomposition approach that generalizes the Bayesian matrix factorization model of \citet{salakhutdinov2008bayesian}, incorporating independent and normally distributed errors.  However, their implementation only provides a point estimate with no uncertainty and does not account for correlation in the residual error structure.

To our knowledge no existing tensor imputation methods perform multiple imputation, in which multiple values are simulated for the missing entries to reflect their uncertainty.  However, this is often critical for applications,  to accurately propagate uncertainty through subsequent analyses after imputation.  As shown in our examples, imputing missing values with a point estimate and subsequently treating the values as fixed can drastically underestimate uncertainty and lead to inaccurate inference.  Thus, we introduce  Bayesian Multiple Imputation for Tensor Arrays (BAMITA), a flexible Bayesian model based on the CP factorization with an efficient MCMC sampling algorithm for  multi-way data with missing values.    Our approach enables valid inference via simulating from the posterior predictive distribution for missing values, accounting for the uncertainty of the imputed elements. Further, our approach incorporates a separable covariance structure on the error terms to account for any correlation structure that may exist on specific tensor modes. We illustrate the advantages of this approach with respect to both imputation accuracy and uncertainty calibration via simulations, and on data from two longitudinal microbiome studies. 

\section{Preliminaries}\label{sec:2}
\subsection{Notation and Background}
Denote $\pmb{\mathscr{X}}\in \mathbb{R}^{I_1\times I_2\times...\times I_N}$ as an $N$-dimensional tensor with the length of each dimension being $I_1$, $I_2,..., I_N$ respectively. Let $\mathbf{a} \circ\mathbf{b}$ denote the outer product of two vectors $\mathbf{a}$ and $\mathbf{b}$. Denote the Kronecker product of matrices $\mathbf{A}$ and $\mathbf{B}$ as $\mathbf{A} \otimes \mathbf{B}$ and the \it{Khatri-Rao} \normalfont product (which is the columnwise Kronecker product) as $\mathbf{A} \odot \mathbf{B}$. Let $\mathbf{X}_{(n)}\in \mathbb{R}^{I_n \times I_1 I_2 ... I_{n-1} I_{n+1}... I_N}$ denote the mode-$n$ matricization of a tensor $\pmb{\mathscr{X}}\in \mathbb{R}^{I_1\times I_2\times ...\times I_N}$, yielding a matrix with $I_n$ rows and  $\prod_{i\neq n}I_i$ columns. Denote the Moore-Penrose pseudoinverse of matrix $\mathbf{A}$ as $\mathbf{A}^{\dagger}$.

For our context, elements of the full tensor $\pmb{\mathscr{X}}$ may be missing. Let $\mathcal{M}=\{(i_1,...,i_N): \mathscr{X}_{i_1,...,i_N} \textnormal{ is missing} \}$ index missing entries, $\pmb{\mathscr{X}}_{\textnormal{missing}}$ give the latent missing values in the tensor $\{\pmb{\mathscr{X}}_{i_1,...,i_N}$ :$ (i_1,...,i_N)\in\mathcal{M}\}$, and $\pmb{\mathscr{X}}_{\textnormal{observed}}$ give observed values in the tensor. Our goal is to infer $\pmb{\mathscr{X}}_{\textnormal{missing}}$ given $\pmb{\mathscr{X}}_{\textnormal{observed}}$. 


\subsection{Tensor CANDECOMP/PARAFAC (CP) Decomposition}
For a tensor $\pmb{\mathscr{X}}\in \mathbb{R}^{I_1\times I_2\times...\times I_N}$, the rank $R$ CP decomposition \citep{carroll1970analysis} factorizes it into a sum of component rank-one tensors that are the outer product of vectors in each dimension,
\begin{equation}
    \pmb{\mathscr{X}} \mathbf \approx \sum_{r=1}^{R} \mathbf{u}_r^{1} \circ \mathbf{u}_r^2\circ... \circ \mathbf{u}_r^N,
\end{equation}
where $\mathbf{u}_r^{1}, \mathbf{u}_r^2,..., \mathbf{u}_r^N$ are vectors with length $I_1,..., I_N$ respectively. For $i=1,...,N$, define the matrix 
\begin{equation}\label{formula:2}
    \mathbf{U}^{i}=
    \begin{bmatrix}
        \mathbf{u}_{1}^i & \mathbf{u}_{2}^i & ... & \mathbf{u}_{R}^i\\
    \end{bmatrix}.
\end{equation}
 Then, we can express the CP decomposition in the following concise way:
\begin{equation*}
    \pmb{\mathscr{X}}\approx \llbracket\mathbf{U^1}, \mathbf{U^2},..., \mathbf{U^N} \rrbracket\equiv \sum_{r=1}^{R} \mathbf{u}_r^{1} \circ \mathbf{u}_r^2\circ... \circ \mathbf{u}_r^N.
\end{equation*}
By definition, the norm of each component is not identifiable, i.e.,
\begin{equation*}
    \llbracket \mathbf{X^1}\Lambda, \mathbf{X^2},..., \mathbf{X^N} \rrbracket=\llbracket \mathbf{X^1}, \mathbf{X^2}\Lambda,..., \mathbf{X^N} \rrbracket=\llbracket \mathbf{X^1}, \mathbf{X^2},..., \mathbf{X^N}\Lambda \rrbracket
\end{equation*}
where $\Lambda=$ diag$(\lambda_1,...,\lambda_R)$. Thus, define the following normalized format where the columns of $\mathbf{U^1}, \mathbf{U^2}$,..., and $\mathbf{U^N}$ have norm 1 and $\pmb{\lambda}$ gives the scale factors: 
\begin{equation}
    \pmb{\mathscr{X}}\approx \llbracket \pmb{\lambda} ; \mathbf{U^1}, \mathbf{U^2},..., \mathbf{U^N} \rrbracket\equiv \sum_{r=1}^{R} \lambda_r \mathbf{u}_r^1 \circ \mathbf{u}_r^2 \circ...\circ \mathbf{u}_r^N.
    \label{scaled_fac}
\end{equation}
We can also write express the factorization in the matricized form
\begin{equation}
    \mathbf{X}_{(i)}=\mathbf{U^i} \Lambda(\mathbf{U^N}\odot...\odot\mathbf{U^{i+1}}\odot\mathbf{U^{i-1}}\odot...\odot\mathbf{U^{1}})^T
\end{equation}
which will be used in our following derivations.

\subsection{Alternating Least Squares (ALS) and EM imputation}
\label{als_and_em}
The alternating least square (ALS) method aims to compute a CP decomposition with $R$ components $\hat{\pmb{\mathscr{X}}}=\sum_{r=1}^{R} \mathbf{u}_r^{1} \circ \mathbf{u}_r^2\circ... \circ \mathbf{u}_r^N$ that best approximates the target tensor $\pmb{\mathscr{X}}\in\mathbb{R}^{I_1\times I_2\times...\times I_N}$ via minimizing the sum of squared residuals:
\begin{equation*}
    \underset{\hat{\pmb{\mathscr{X}}}}{\text{min}}||\pmb{\mathscr{X}}-\hat{\pmb{\mathscr{X}}}||^2. \label{sumofsquares}
\end{equation*}
The ALS algorithm iteratively achieves this by solving the least square problem of one component ($\hat{\mathbf{U}}^1$, for example) with the other components fixed, which is
\begin{equation*}
    \begin{split}
        \hat{\mathbf{U}}^1&=\underset{{\mathbf{U}^1}}{\text{argmin}}||\pmb{\mathscr{X}}-\llbracket \pmb{\lambda} ; \mathbf{U^1}, \mathbf{U^2},..., \mathbf{U^N} \rrbracket||_{F}\\
        &=\underset{{\mathbf{U}^1}}{\text{argmin}}||\mathbf{X}_{(1)}-\mathbf{U}^1 \Lambda(\mathbf{U}^N\odot...\odot\mathbf{U}^2)^T||_{F}.
    \end{split}
\end{equation*}
Solving the least square equation, we have 
\begin{equation*}
    \begin{split}
        \hat{\mathbf{U}}^1&=\mathbf{X}_{(1)}[(\mathbf{U}^N\odot...\odot\mathbf{U}^2)^T]^{\dagger}\\
        &=\mathbf{X}_{(1)}(\mathbf{U}^N\odot...\odot\mathbf{U}^2)[(\mathbf{U}^N)^T\mathbf{U}^N*...*(\mathbf{U}^2)^T\mathbf{U}^2]^{\dagger}.
    \end{split}
\end{equation*}
Where $*$ is the Hadamard product which is the elementwise matrix product for two matrices with the same dimensions.
We can normalize each column of $\hat{\mathbf{U}}^1=\begin{bmatrix}
            \hat{\mathbf{u}}^1_{1} & \hat{\mathbf{u}}^1_{2} & ... & \hat{\mathbf{u}}^1_{R}\\
        \end{bmatrix}$ to calculate $\mathbf{U}^1$ and $\Lambda$ where $\lambda_r=||\hat{\mathbf{u}}^1_{r}||$ and $\mathbf{u}^1_r=\hat{\mathbf{u}}^1_r/\lambda_r$.
Then, we fix $\mathbf{U}^1,\mathbf{U}^3...,\mathbf{U}^N$ to compute $\mathbf{U}^2$ in a similar way. In this way, $\mathbf{U}^1...,\mathbf{U}^N$ are iteratively updated until the algorithm converges (i.e., the calculated matrix is very close to the former matrix) or it attains the maximum number of iterations.

A common frequentist approach to impute missing data for tensors is an expectation-maximization (EM) algorithm, in which missing entries are iteratively updated via ALS or a similar procedure \citep{acar2011scalable}. Missing data are first initialized, e.g.,  as $\pmb{\mathscr{X}}_{i_1,...,i_N} =0$ for $(i_1,...,i_N)\in\mathcal{M}$ if the tensor is centered. Then, the solution to \eqref{sumofsquares} is determined for the full tensor $\pmb{\mathscr{X}}$, the missing entries are updated as their corresponding elements in $\pmb{\hat{\mathscr{X}}}$, and the process is repeated until convergence.

\section{Bayesian Tensor Imputation Algorithm}
In this section, we describe BAMITA, a Bayesian approach for the CP decomposition under different assumptions on the variance structure. We first derive the Bayesian MCMC sampling procedure under the assumption that the error (noise) term is independent and normally distributed. We then introduce another model where the error term is assumed to be normally distributed with a separable covariance structure. 
\subsection{Independent error}
\subsubsection{The model}
 First, follow the notation in Section \ref{sec:2}, assume the following model based in the CP decomposition with rank $R$: 
\begin{equation}\label{equ: model}
    \pmb{\mathscr{X}}=\llbracket {\mathbf{U}^{1}}, \mathbf{U}^{2},..., \mathbf{U}^{N} \rrbracket + \pmb{\mathscr{E}},
\end{equation}
where $\pmb{\mathscr{X}}\in \mathbb{R}^{I_1\times I_2\times...\times I_N}$, $\mathbf{U}^{1},...,\mathbf{U}^{N}$ defined in (\ref{formula:2}) with $R$ components (i.e., rank $R$), and $\pmb{\mathscr{E}}$ is a tensor of the error terms with the same dimensions. In the rest of this section, we treat $R$ as fixed. In practice, the number of components $R$ can be selected through cross-validation. Here, we assume the error terms are independent and identically distributed from a normal distribution with mean $0$ and variance $\sigma^2$. As our goal is to have a non-informative prior that can be used in a variety of data situations, we consider an (improper) conjugate Jeffreys prior for all ${\mathbf{U}^{1}}, \mathbf{U}^{2},..., \mathbf{U}^{N}$ and $\sigma^2$, with,
\begin{equation*}
    p(\mathbf{U}^{n})\propto 1
\end{equation*}
for $n=1,...,N$ and 
\begin{equation*}
    p(\sigma^2)\propto \frac{1}{\sigma^2}.
\end{equation*}
Although the prior we impose is improper, the corresponding posterior will be proper as we will demonstrate that all the conditional distributions for the posterior are proper. Moreover, the model is invariant to the relative scale of the ${\mathbf{U}^{i}}$; one can scale the individual factors as in \eqref{scaled_fac}, but this will not affect posterior inference for the low-rank structure and missing data, which is our primary focus.  
\color{black}

\subsubsection{Gibbs Sampler with full data}
Here we give full conditionals for the Gibbs sampling procedure for fully observed data. Note that, given $\mathbf{U}^{2},..., \mathbf{U}^{N}$ and $\sigma^2$, the model can be expressed as:
\begin{equation}
    \X_{(1)}=\mathbf{U}^{1}(\mathbf{U}^{N}\odot \mathbf{U}^{(N-1)}\odot...\odot \mathbf{U}^{2})^{T}+\mathbf{E}_{(1)}
\end{equation}
where $\X_{(1)}\in\mathbb{R}^{I_1\times I_2...I_N}$ is the mode-1 matricization of the tensor $\hat{\pmb{\mathscr{X}}}=\llbracket {\mathbf{U}^{1}}, \mathbf{U}^{2},..., \mathbf{U}^{N} \rrbracket$ and $\mathbf{E}_{(1)}$ is the mode-1 matricization of the error tensor $\pmb{\mathscr{E}}$ with the same dimension as $\X_{(1)}$. \ Let $\A_{(1)}=(\mathbf{U}^{N}\odot \mathbf{U}^{(N-1)}\odot...\odot \mathbf{U}^{2})$ where the subscript $1$ indicate that it is the Khatri-Rao product without matrix $\boldsymbol{U}^1$ \color{black}. We partition $\X_{(1)}$, $\mathbf{U}^{1}$, and $\E_{(1)}$ along their row, i.e., 
\begin{equation*}
    \X_{(1)}=
    \begin{bmatrix}
        (\x^1_{1\cdot})^T & (\x^1_{2\cdot})^T & ... & (\x^1_{I_1\cdot})^T\\
    \end{bmatrix}^T
\end{equation*}
and,
\begin{equation*}
    \mathbf{U}^{1}=
    \begin{bmatrix}
        (\mathbf{u}^1_{1\cdot})^T & (\mathbf{u}^1_{2\cdot})^T & ... & (\mathbf{u}^1_{I_1\cdot})^T\\
    \end{bmatrix}^T
\end{equation*}
and, 
\begin{equation*}
    \E_{(1)}=
    \begin{bmatrix}
        (\mathbf{\epsilon}^1_{1\cdot})^T & (\mathbf{\epsilon}^1_{2\cdot})^T & ... & (\mathbf{\epsilon}^1_{I_1\cdot})^T\\
    \end{bmatrix}^T.
\end{equation*}
Since each element in the error term is assumed to be independent and normally distributed, we have the following Bayesian linear regression model
\begin{equation}
    \x^1_{i\cdot} = \mathbf{u}^1_{i\cdot} \A_{(1)} + \mathbf{\epsilon}^1_{i\cdot}
\end{equation}
for $i = 1,..., I_1$ where $\mathbf{\epsilon}^1_{i\cdot}$ be i.i.d. normal distributed with mean 0 and variance $\sigma^2$. Then, with the Jeffreys priors $p(\mathbf{u}^1_{i\cdot})\propto1$ and $p(\sigma^2)\propto\frac{1}{\sigma^2}$, the posterior distribution for $\mathbf{u}^1_{i\cdot}$ is a normal distribution with mean $(\A_{(1)}^T\A_{(1)})^{-1}\A_{(1)}^T\x^1_{i\cdot}$ and covariance matrix $\sigma^2(\A_{(1)}^T\A_{(1)})^{-1}$. Therefore, we have the following Gibbs sampling algorithm for fully observed data: 


\begin{itemize}
    \item Given $\mathbf{U}^{2},..., \mathbf{U}^{N}$ and $\sigma^2$, draw 
    \begin{equation*}
    \mathbf{U}^{1}=
    \begin{bmatrix}
        (\mathbf{u}^1_{1\cdot})^T & (\mathbf{u}^1_{2\cdot})^T & ... & (\mathbf{u}^1_{I_1\cdot})^T\\
    \end{bmatrix}^T
    \end{equation*} with 
    \begin{equation*}
        \mathbf{u}^1_{i\cdot} \sim N((\A_{(1)}^T\A_{(1)})^{-1}\A_{(1)}^T\x^1_{i\cdot}, \sigma^2(\A_{(1)}^T\A_{(1)})^{-1})
    \end{equation*} 
    for  $i = 1,..., I_1$ where $\A_{(1)}=(\mathbf{U}^{N}\odot \mathbf{U}^{(N-1)}\odot...\odot \mathbf{U}^{2})$.
    
    \item Draw $\mathbf{U}^{2},..., \mathbf{U}^{N}$ similarly:
    \begin{equation*}
    \mathbf{U}^{n}=
    \begin{bmatrix}
        (\mathbf{u}^n_{1\cdot})^T & (\mathbf{u}^n_{2\cdot})^T & ... & (\mathbf{u}^n_{I_n\cdot})^T\\
    \end{bmatrix}^T
    \end{equation*} with 
    \begin{equation*}
        \mathbf{u}^n_{i\cdot} \sim N((\A_{(n)}^T\A_{(n)})^{-1}\A_{(n)}^T\x^1_{i\cdot}, \sigma^2(\A_{(n)}^T\A_{(n)})^{-1})
    \end{equation*} 
    for  $i = 1,..., I_n$ where $\A_{(n)}=(\mathbf{U}^{N}\odot...\odot \mathbf{U}^{(n+1)}\odot\mathbf{U}^{(n-1)}\odot...\odot \mathbf{U}^{1})$.

    \item Given $\mathbf{U}^{1},..., \mathbf{U}^{N}$, draw $\sigma^2$ via its inverse-gamma full conditional IG$(\alpha=(I_1\cdot ...\cdot I_N)/2, \beta=(||\pmb{\mathscr{X}}-\hat{\pmb{\mathscr{X}}}||_{F})/2)$ where $\hat{\pmb{\mathscr{X}}}$ is calculated using the simulated $\mathbf{U}^{1},...,\mathbf{U}^{N}$ from the previous steps, and $||\cdot||_F$ be the Frobenius norm.
\end{itemize}

\subsubsection{Missing data imputation}
We propose to impute the missing entries $\pmb{\mathscr{X}}_{i_1,...,i_N}, (i_1,...,i_N)\in\mathcal{M}$ with simulated values from their posterior predictive distribution $p(\textnormal{Vec}(\pmb{\mathscr{X}}_{\textnormal{missing}}) | \textnormal{Vec}(\pmb{\mathscr{X}}_{\textnormal{observed}}))$ for each MCMC iteration.  Note that, if we assume the error terms are independently distributed, the conditional distribution will simply be determined by the posterior distribution for the error variance and low-rank structure  for the  missing entries. The missing entries will be imputed according to their posterior sampling.
Given a partially observed tensor $\pmb{\mathscr{X}}$ and number of modes $R$, our Bayesian multiple imputation algorithm for normal i.i.d error is given in Algorithm~\ref{alg1}.

\begin{algorithm}
	\caption{Bayesian multiple imputation with normal i.i.d error} 
 \label{alg1}
	\begin{algorithmic}[1]
        \State Impute the missing elements as $\pmb{\mathscr{X}}_{i_1,...,i_N}=0, (i_1,...,i_N)\in \mathcal{M}$.
	  \State Set the initial value of $(\hat{\mathbf{U}}^{1})^0$,..., $(\hat{\mathbf{U}}^{N})^0$ as either randomly generated number or the result of the frequentist EM algorithm. Set the initial value of $(\hat{\sigma}^2)^{0}\sim$ IG$(\alpha=(I_1\cdot ...\cdot I_N)/2, \beta=(||\pmb{\mathscr{X}}-\hat{\pmb{\mathscr{X}}}||_{F})/2)$ where $\hat{\pmb{\mathscr{X}}}$ is calculated using the initial value of $(\hat{\mathbf{U}}^{1})^0$,..., $(\hat{\mathbf{U}}^{N})^0$.

        \For {$r=1,...,B$-th MCMC iteration}
            \State Sample $(\hat{\mathbf{U}}^{1})^r$,..., $(\hat{\mathbf{U}}^{N})^r$, and $(\hat{\sigma}^2)^{r}$ with the previous posterior distributions.
            \State Calculate the underlying structure of $\pmb{\mathscr{X}}$ as $\Tilde{\pmb{\mathscr{X}}}^r = \llbracket {(\mathbf{U}^{1})^r},..., (\mathbf{U}^{N})^r \rrbracket $
            \State Calculate $\hat{\pmb{\mathscr{X}}}^r$ where missing elements $\pmb{\mathscr{X}}_{i_1...i_N}, (i_1...i_N)\in\mathcal{M}$ are imputed following $N(\Tilde{\pmb{\mathscr{X}}}_{i_1,...,i_N}^r, (\hat{\sigma}^2)^{r})$
        \EndFor		
        \State Impute the missing elements $\pmb{\mathscr{X}}_{i_1...i_N}, (i_1,...,i_N)\in\mathcal{M}$ as mean value of $\{\Tilde{\pmb{\mathscr{X}}}_{i_1,...,i_N}^b,...,\Tilde{\pmb{\mathscr{X}}}_{i_1,...,i_N}^B\}$ where $b$ be the burn-in value of the Gibbs sampler.

        

	\end{algorithmic} 
\end{algorithm}

\
In practice, the rank $R$ can be determined by cross-validation. In Algorithm~\ref{alg2}, we describe a general cross-validation procedure for choosing $R$ using the mean squared error (MSE) of the imputed held-out elements. 

\begin{algorithm}
	\caption{Selecting number of the component with cross validation} 
 \label{alg2}
	\begin{algorithmic}[1]
        
        \State Randomly divide the observed elements into $K$ equal folds with index $\mathcal{F}_1,...,\mathcal{F}_K$.
        \For {$r=1,...,R$ number of components}
        \For {$k=1,...,K$-th fold}
        \State Hold out the elements in the $k$-th fold $\pmb{\mathscr{X}}_{i_1,...,i_N}=0, (i_1,...,i_N)\in \mathcal{F}_k$ as missing.
	  \State Run the corresponding Bayesian imputation algorithm (Algorithm \ref{alg1} for independent error) and get the imputed value $\hat{\pmb{\mathscr{X}}}_{i_1,...,i_N}, (i_1,...,i_N)\in \mathcal{F}_k$ for the held-out data.
        \State Calculate the mean squared error (MSE) for the imputation over the held-out data. $\delta_k=\sum_{(i_1,...,i_N)\in \mathcal{F}_k} (\hat{\pmb{\mathscr{X}}}_{i_1,...,i_N}-\pmb{\mathscr{X}}_{i_1,...,i_N})^2$.
        
        \EndFor	
        
        \State Calculate the average MSE with number of component equals $r$ as $\Bar{\delta}^r=\frac{1}{K}\sum_{k=1}^K \delta_k$.
        
        \EndFor		
        \State Select the optimal number of components with the smallest average MSE.

        

	\end{algorithmic} 
\end{algorithm}

\color{black}

\subsection{Correlated error}
\subsubsection{The model}
For many tensor applications, data have a residual correlation along some of the modes that is not efficiently captured by the low-rank decomposition. For example, in our longitudinal infant gut microbiome data, observations are assumed to be correlated across time intervals for the same participants, and abundances are correlated across the taxa. Therefore, instead of assuming the errors are independent, here we assume the errors are normally distributed but correlated. For simplicity, we assume the error tensor in (\ref{equ: model}) has the separable covariance structure described by \citet{hoff2011separable}:
\begin{equation}
    \pmb{\mathscr{E}}\sim N_{I_1\times I_2\times...\times I_N}(\mathbf{0},\Sigma_1,...,\Sigma_N),
\end{equation}
where the tensor normal distribution $N_{I_1\times I_2\times...\times I_N}(\mathbf{0},\Sigma_1,...,\Sigma_N)$ is defined according to its $n$-th mode unfolding $\pmb{\mathscr{E}}_{(n)}$ which follows a matrix normal distribution:
\begin{equation*}
    \pmb{\mathscr{E}}_{(n)}\sim N_{I_n\times I_{-n}}(\mathbf{0},\Sigma_n,\Sigma_{-n})
\end{equation*}
where $I_{-n}=\prod_{i=1,...,N;i\neq n} I_i$ and $\Sigma_{-n}=\Sigma_1\otimes\Sigma_2\otimes...\Sigma_{n-1}\otimes\Sigma_{n+1}\otimes...\Sigma_N$. \color{black} The vectorization of the tensor then follows a multivariate normal distribution:
\begin{equation*}
    \text{Vec}(\pmb{\mathscr{E}})\sim N_{I_1\cdot I_2...\cdot I_N}(\mathbf{0},\Sigma_N \otimes \Sigma_{N-1}...\otimes \Sigma_1).
\end{equation*}

Compared with estimating the entire unrestricted covariance matrix $\textnormal{Cov}(\textnormal{Vec}(\pmb{\mathscr{E}}))$ which can be unrealistic due to its potentially high dimension, using a separable covariance structure provides a more stable and parsimonious way of modeling the correlation along different modes. Moreover, it is also more interpretable for the covariance matrices $\Sigma_i$ for each mode, e.g., a covariance for time points and a covariance for taxa.

Similar to the independent error, we use a non-informative prior to make our algorithm broadly accommodating, with the uniform prior on each $\mathbf{U}^{1},...,\mathbf{U}^{N}$ 
\begin{equation*}
    \mathbf{U}^{n} \propto 1
\end{equation*}
and inverse-Wishart prior on $\Sigma_1,...,\Sigma_N$,
\begin{equation*}
    P(\Sigma_n)\sim \text{IW}(\text{Diag}(I_n),(I_n+2)).
\end{equation*}
In practice, the set of inverse-Wishart priors for each dimension depends on the specific scientific context of the application. In our implementation, the first dimension is considered the sample dimension and samples are independent. Thus, we set $\Sigma_1$ to the identity and use a inverse-Wishart prior on the other dimensions.

\subsubsection{Bayesian Gibbs Sampler for full data}
Similar to the independent case, our model facilitates a conjugate MCMC sampling procedure for fully observed data:

\begin{itemize}
    \item Given $\mathbf{U}^{2},..., \mathbf{U}^{N}$ and $\Sigma_2,...,\Sigma_N$, sample $\Sigma_1$ with
    \begin{equation*}
        \Sigma_1|\mathbf{U}^{1},\mathbf{U}^{2},..., \mathbf{U}^{N},\Sigma_2,...,\Sigma_N\sim \text{IW}(S_n, I_n+2+I_{-n} )
    \end{equation*}
    where $S_n=\text{Diag}(I_n)+((\Tilde{\A}^T\Tilde{\A})^{-1}\Tilde{\A}^T\Tilde{\X}_{(1)}\Tilde{\A}-\Tilde{\X}_{(1)})^T((\Tilde{\A}^T\Tilde{\A})^{-1}\Tilde{\A}^T\Tilde{\X}_{(1)}\Tilde{\A}-\Tilde{\X}_{(1)})$, $\Tilde{\A}=\A\Sigma_{-1}^{-1/2}$, and $\Tilde{\X}_{(1)}=\X_{(1)} \Sigma_{-1}^{-1/2}$.
    \item Given $\Sigma_1$ and $\mathbf{U}^{2},..., \mathbf{U}^{N}$ and $\Sigma_2,...,\Sigma_N$, sample $\mathbf{U}^{1}$ with
    \begin{equation*}
         \mathbf{U}^{1}|\mathbf{U}^{2},..., \mathbf{U}^{N},\Sigma_1,...,\Sigma_N \sim \text{MatrixNormal}_{}((\Tilde{\A}^T\Tilde{\A})^{-1}\Tilde{\A}^T\Tilde{\X}_{(1)}, (\Tilde{\A}^T\Tilde{\A})^{-1}, \Sigma_1)
    \end{equation*}
    
    \item Sample $\mathbf{U}^{2},\Sigma_2 ;...; \mathbf{U}^{N},\Sigma_N$ in a similar way. 
\end{itemize}
\subsubsection{Missing data imputation}
The multiple imputation algorithm for $\pmb{\mathscr{X}}_{missing}$ given $\pmb{\mathscr{X}}_{observed}$ proceeds somewhat differently for a separable covariance structure than that under independence in Algorithm \ref{alg1}, because the observed values provide additional information to inform missing elements beyond the low rank signal.  Let $\boldsymbol{\mu_m}$ and $\boldsymbol{\mu_o}$ be the vectorized mean of the missing and observed entries, respectively, given by the low-rank structure (\ref{formula:2}). Under the multivariate normal error term with separable covariance structure, the conditional posterior predictive distribution of the missing entries will be 
\begin{equation}\label{post_condi}
    p(\textnormal{Vec}(\pmb{\mathscr{X}}_{\textnormal{missing}}) | \textnormal{Vec}(\pmb{\mathscr{X}}_{\textnormal{observed}}))\sim N(\boldsymbol{\mu}_{\textnormal{missing}|\textnormal{observed}}, \Sigma_{\textnormal{missing}|\textnormal{observed}})
\end{equation}
where
\begin{equation*}
    \boldsymbol{\mu}_{\textnormal{missing}|\textnormal{observed}}=\boldsymbol{\mu_m}+\Sigma_{12}(\Sigma_{22})^{-1}(\pmb{\mathscr{X}}_{observed}-\boldsymbol{\mu_o})
\end{equation*}
and
\begin{equation*}
    \Sigma_{\textnormal{missing}|\textnormal{observed}}=\Sigma_{11}-\Sigma_{12}(\Sigma_{22})^{-1}\Sigma_{21}.
\end{equation*}
Here, $\Sigma_{11}, \Sigma_{12}, \Sigma_{21}, \Sigma_{22}$ partition the full covariance matrix of the vectorized tensor according to the missing indices (i.e., $\Sigma_{11}=\textnormal{Cov}(\boldsymbol{\mu_m},\boldsymbol{\mu_m})$ and $\Sigma_{21}=\textnormal{Cov}(\boldsymbol{\mu_o},\boldsymbol{\mu_m})$, etc). 

\begin{algorithm}
	\caption{Bayesian multiple imputation with correlated error} 
 \label{alg3}
	\begin{algorithmic}[1]
        \State Impute the missing elements in $\pmb{\mathscr{X}}$ as 0.
	  \State Set the initial value of $(\hat{\mathbf{U}}^{1})^0$,..., $(\hat{\mathbf{U}}^{N})^0$ as either randomly generated number or the result of the frequentist EM algorithm. Set the initial value of $(\hat{\Sigma}_n)^0$ as standard diagonal matrix with dimension $I_n$ for $n=1,...,N$.

        \For {$r=1,...,B$-th MCMC iteration}
            \State Sample $(\hat{\mathbf{U}}^{1})^r,(\hat{\Sigma}_1)^r ;...; (\hat{\mathbf{U}}^{N})^r,(\hat{\Sigma}_N)^r$ with the previous posterior distributions.
            \State Calculate the underlying structure of $\pmb{\mathscr{X}}$ as $\Tilde{\pmb{\mathscr{X}}}^r = \llbracket {(\mathbf{U}^{1})^r},..., (\mathbf{U}^{N})^r \rrbracket $
            \State Partition the mean Vec$(\Tilde{\pmb{\mathscr{X}}}^r)$ into Vec$(\Tilde{\pmb{\mathscr{X}}}^r)_o$ and Vec$(\Tilde{\pmb{\mathscr{X}}}^r)_m$. Partition $\Tilde{\Sigma}^r=\hat{\Sigma}_N^r \otimes \hat{\Sigma}_{N-1}^r...\otimes \hat{\Sigma}_1^r$ into $\Sigma_{11}^r$, $\Sigma_{12}^r$, $\Sigma_{21}^r$, and $\Sigma_{22}^r$ according to the corresponding missingness of the elements.
            
            \State Simulate $\hat{\pmb{\mathscr{X}}}^r_{\textnormal{missing}}$ according to the posterior predictive distribution described in \ref{post_condi} and get $\hat{\pmb{\mathscr{X}}}^r$ where the missing entries are imputed with $\hat{\pmb{\mathscr{X}}}^r_{\textnormal{missing}}$.
        \EndFor		
        \State Impute the missing entries $\pmb{\mathscr{X}}_{i_1...i_N}$ as mean value of $\{\Tilde{\pmb{\mathscr{X}}}_{i_1,...,i_N}^b,...,\Tilde{\pmb{\mathscr{X}}}_{i_1,...,i_N}^B\}$ where $b$ be the burn-in value of the Gibbs sampler.
	\end{algorithmic} 
\end{algorithm}

Given a observed tensor $\pmb{\mathscr{X}}_{observed}$ and number of modes $N$, we now describe an algorithm to simulate from the posterior predictive distribution, $p(\pmb{\mathscr{X}}_{missing} | \pmb{\mathscr{X}}_{observed})$, for missing entries. Our Bayesian multiple imputation algorithm for error term with separable covariance structure is described in Algorithm~\ref{alg3}.

Note that, in our implementation, to be consistent with our data application, the first dimension is considered the sample dimension and samples are independent. Thus, we set $\Sigma_1$ to the identity and use an inverse-Wishart prior on $\Sigma_2$,...,$\Sigma_N$. In practice, whether or not the full covariance is modeled for a given dimension can depend on the context of the application.

\section{Simulation}

To evaluate the performance of our proposed Bayesian tensor imputation methods, we conducted a series of simulation experiments. Since we propose two Bayesian imputation methods for multiway data, we aim to use the simulation experiments to evaluate:

1. The performance of the Bayesian independent imputation algorithm in terms of rank selection under the cross-validation.

2. The performance of the Bayesian multiple imputation algorithms with independent or correlated error in terms of the imputed data entries.

3. The performance of the Bayesian multiple imputation algorithms with independent or correlated error in terms of fiber-wise imputation and inferring uncertainty for functions of the fibers.

For each simulation condition, we run 100 experiments with two independent MCMC chains. All the results are calculated as the median of the converged experiments of the Bayesian algorithm, where the convergence is evaluated using a composite version of the scale reduction factor defined as:
\begin{equation*}
    \textnormal{SRF} = \frac{2\sum_{i=1}^{n_1+n_2}(X_{i}-\Bar{X})^2/(n_1+n_2-1)}{\sum_{i=1}^{n_1}(X_{1i}-\Bar{X}_1)^2/(n_1-1)+\sum_{i=1}^{n_2}(X_{2i}-\Bar{X}_2)^2/(n_2-1)}
\end{equation*}
where $\{X_{11}, X_{12},..., X_{1n_1}\}$ and $\{X_{21}, X_{22},..., X_{2n_2}\}$ be the posterior samples of the two separate chains and $\{X_{1}, X_{2},..., X_{n_1+n_2}\}$ is the combination of those chains.


\subsection{Simulation study 1: Rank selection and independent error}
For our first study,  simulated data $\pmb{\mathscr{X}}$ of dimension $I_1\times I_2\times I_3$ is generated following (\ref{equ: model}), 
  where $\llbracket {\mathbf{U}^{(1)}}, \mathbf{U}^{(2)}, \mathbf{U}^{(3)} \rrbracket$ has a rank-$3$ underlying structure and $\pmb{\mathscr{E}}$ has independent normal error with mean $0$ and variance $1$. Each $ {\mathbf{U}^{(i)}}$, $i=1,2,3$ is a matrix of dimension $I_i\times 3$ whose elements are drawn from a standard normal distribution. To demonstrate the performance of our algorithm in different scales of dimensions, we considered three scenarios: $(10 \times 10\times 10)$, $(20 \times 20\times 20)$, and a higher-dimensional imbalanced scenario $(10 \times 100\times 1000)$. Different missing patterns and proportions of the tensor elements are also considered. For an entry-wise missing scenario, each element of the tensor is randomly set to be missing with the probability of 0.2, 0.5, or 0.7 (yielding missing proportions of 20\%, 50\%, or 70\%, respectively). For a fiber-wise missing scenario, the elements for the entire fiber of the third mode are randomly set to be missing with the corresponding probability. 
  
We run our algorithms with the assigned number of components (rank) equal to $1,2,3,4,$ or $5$ (the true rank is 3) and select the rank according to cross-validation. The validation set is composed of the elements randomly drawn from the observed elements (i.e., not the elements which are set to be missing) of the simulated tensor data. The rank is selected based on the mean squared error (MSE) over the validation set (25\% of all the observed elements). For the fiber-wise missing condition, the validation set is also randomly assigned for the entire fiber to be consistent. We evaluate the performance of our Bayesian independent imputation algorithm based on the mean squared error (MSE) and the coverage rate of the $95\%$ credible interval with the true rank and the selected rank. \ The upper
and lower bounds for the 95\% credible interval for each element are calculated using the 0.025 and 0.975 quantiles of the MCMC samples.
The MSE and coverage are calculated as median values over all the missing elements. \color{black}


We compare the performance of our Bayesian multiple imputation with independent error against the frequentist EM algorithm with the true rank described in Section \ref{als_and_em}. Results are presented in Table~\ref{Table1}. The MSE for the true rank and the selected rank are very close across conditions, which indicates the reasonable performance of rank selection via cross-validation. In fact, the cross-validation selects the true rank in most of the experiments except for the scenarios of $10\times 10\times 10$ dimension and $70\%$ missingness.

Coverage rates for credible intervals are all approximately $95\%$, so uncertainty is correctly inferred. Morevoer, our Bayesian independent imputation method performs better than the EM algorithm in terms of the MSE in most of the scenarios. When the dimension of the tensor is $(10\times 100\times 1000)$, the two algorithms have similar performance. This matches our expectation since when the total sample size is large enough (for the element-wise missing), the results solved by the frequentist EM algorithm are generally similar to the posterior mean for the Bayesian model with flat priors and independent error.  MCMC convergence for the fixed number of iterations is generally achieved, but less so for scenarios with 70\% missing fibers.  

\begin{center}
\begin{table}[!h]
\caption{Simulation results for study 1: rank selection and independent error. Results are presented as median values over all the missing elements. The best performance in each setting was marked in boldface.\label{Table1}}
\resizebox{\textwidth}{!}{
\begin{tabular}{cccccccccccccc}
\hline
\multicolumn{1}{c}{\textbf{Missing}} & 
\multicolumn{1}{c}{\textbf{Tensor}} & 
\multicolumn{1}{c}{\textbf{Missing}} & &
\multicolumn{4}{c}{\textbf{BAMITA independent}} &  & 
\multicolumn{1}{c}{\textbf{EM Algorithm}} &
\multicolumn{1}{c}{\textbf{True}}&
\multicolumn{1}{c}{\textbf{Selected}}&
\multicolumn{1}{c}{\textbf{Overall}}\\ 
\cline{5-8} 
\multicolumn{1}{c}{\textbf{pattern}} & 
\multicolumn{1}{c}{\textbf{dimension}} & 
\multicolumn{1}{c}{\textbf{proportion}} & &
\multicolumn{2}{c}{\textbf{True Rank}} &
\multicolumn{2}{c}{\textbf{Selected Rank}}& &
\multicolumn{1}{c}{\textbf{True Rank}}&
\multicolumn{1}{c}{\textbf{Converged}}&
\multicolumn{1}{c}{\textbf{Converged}}&
\multicolumn{1}{c}{\textbf{Converged}}&
 \\
\cline{5-8} 
&  &  & &  
\textbf{MSE} & 
\textbf{Coverage(\%)}  & 
\textbf{MSE} & 
\textbf{Coverage(\%)} & & 
\textbf{MSE}  & 
\textbf{\%}  & 
\textbf{\%}  & 
\textbf{\%} \\ 
\hline

Entry Missing & (10$\times$ 10$\times$ 10) & 20\% 
& & 0.399 & \bf{94.8} & \bf{0.327} & 94.7&  & 0.363 & 99 & 100& 99 \\
 & & 50\% & & 0.566 & \bf{95.2} & \bf{0.508} & 92.6&  & 0.718 & 96 & 100& 96 \\
  &  & 70\% & & 1.323 & \bf{93.0} & \bf{0.869} & 86.0&  & 1.136 & 90 & 91& 85 \\
   \\
 & (20$\times$ 20$\times$ 20) 
 & 20\% & & 0.269 & 94.4 & \bf{0.267} & \bf{94.5}&  & 0.266 & 99& 100& 99 \\
 & & 50\% & & 0.306 & 94.5 & \bf{0.302} & \bf{94.7}&  & 0.326 & 94& 99& 94  \\
  &  & 70\% & & 0.357 & 94.0 & \bf{0.335} & \bf{95.0}&  & 0.522 & 76& 100& 76 \\
   \\
 & (10$\times$ 100$\times$ 1000) 
 & 20\% & & \bf{0.257} & {94.4} & \bf{0.257} & \bf{94.4}&  & 0.259 & 97 & 97& 97 \\
 & & 50\% & & \bf{0.271} & {94.4} & \bf{0.271} & \bf{94.4}&  & 0.272 & 98 & 98& 98  \\
  &  & 70\% & & \bf{0.256} & {94.3} & \bf{0.256} & \bf{94.3}&  & 0.287 & 92 & 92& 91 \\
   \\
  
Fiber Missing & (10$\times$ 10$\times$ 10) 
& 20\% & & \bfseries{0.375} & \bfseries{95.3} & \bfseries{0.375} &\bfseries{95.3}&  & 0.438 & 100& 100& 100  \\
 & & 50\% & & 0.523 & \bfseries{93.1} & \bfseries{0.478} & 92.8&  & 0.876 & 88 & 89& 85 \\
  &  & 70\% & & 0.771 & \bfseries{87.3} & \bfseries{0.641} & 83.8&  & 1.128 & 34 & 43& 23 \\
   \\
 & (20$\times$ 20$\times$ 20) 
 & 20\% & & \bfseries{0.285} & 94.8 & \bfseries{0.285} & 94.8&  & 0.279 & 100& 100& 100  \\
 & & 50\% & & 0.300 & 94.5 & \bfseries{0.298} & \bfseries{94.8}&  & 0.371 & 94& 100& 94  \\
  &  & 70\% & & 0.335 & 90.9 & \bfseries{0.335} & \bfseries{93.1}&  & 0.720 & 65 & 81& 65 \\
   \\
 & (10$\times$ 100$\times$ 1000) 
 & 20\% & & \bfseries{0.251} & \bfseries{94.4} & \bfseries{0.251} & \bfseries{94.4}&  & 0.265 & 100 & 100& 100 \\
 & & 50\% & & \bfseries{0.254} & \bfseries{93.2} & \bfseries{0.254} & \bfseries{93.2}&  & 0.346 & 98& 98& 97   \\
  &  & 70\% & & 0.260 & \bfseries{87.6} & \bfseries{0.259} & \bfseries{87.6}&  & 0.570 & 52  & 48& 47\\
   \\
   \hline
\end{tabular}}
\end{table}
\end{center}

\subsection{Simulation study 2: Tensor imputation and correlated error}
The second simulation study evaluates the performance of our Bayesian imputation algorithm with correlated error, in terms of the imputed entries or fibers in the tensor. Similar to the first simulation, we generate $\pmb{\mathscr{X}}$ according to (\ref{equ: model}) with $\llbracket {\mathbf{U}^{(1)}}, \mathbf{U}^{(2)}, \mathbf{U}^{(3)} \rrbracket$ as rank-$3$ underlying structure. To better mimic the condition of our application data, we simulate $\pmb{\mathscr{X}}$ with dimensions $10\times 10\times 10$, $20\times 20\times 20$, and $65\times 168\times 6$. The error terms $\pmb{\mathscr{E}}$ are generated with separable covariance structure:
  \begin{equation*}
      \pmb{\mathscr{E}}\sim N_{I_1\times I_2\times I_3}(\mathbf{0},\Sigma_1,(\Sigma_2\Sigma_2)^T,(\Sigma_3\Sigma_3^T))
  \end{equation*}
where $\Sigma_i, i=2,3$ are the $I_2 \times I_2$ and $I_3\times I_3$ covariance matrix with diagonal elements being $0.9$ and other elements randomly set to be either $0.3$ or $-0.3$ with equal probability. Following our applications, $\Sigma_1$ is set to be the $I_1\times I_1$ diagonal matrix (i.e., there is no correlation structure among the first dimension) with diagonal elements being $0.5$. The different missing patterns and proportions of the tensor elements are the same as in the first simulation study. In this simulation study the rank is set to be $3$ \color{black} and the performance of each algorithm is directly evaluated over the missing elements.

The missing value is imputed using the posterior mean, and we calculate the mean squared error (MSE) and the coverage rate for the $95\%$ credible interval of the missing elements for the two Bayesian methods. For the frequentist EM method, we calculate the MSE for the missing elements as a comparison.  {We also compare with the missForest method \citep{stekhoven2012missforest} applied to the matricized data along the first mode, as a general approach that does not assume low-rank tensor structure. } 

The results can be seen in Table ~\ref{Table2}. For the 70\% fiber-wise missing with a dimension of $(10\times 10\times 10)$, we do not include the results as only 4\% of the total experiments converged. We can see that the Bayesian correlated imputation algorithm outperforms the Bayesian independent imputation algorithm and the frequentist EM algorithm in most of the scenarios in terms of the MSE except for the 70\% fiber-wise missing case with dimension $(65\times168\times6)$. The coverage of the correlated algorithm is also comparable to the independent algorithm, and both have coverage close to the nominal rate of 95\%.


\begin{center}
\begin{table}[]
\
\caption{Simulation results for study 2: tensor imputation with correlated error. Results are presented as median values over all the missing elements.\label{Table2}}
\resizebox{\textwidth}{!}{
\begin{tabular}{cccccccccccccccc}

\hline
\multicolumn{1}{c}{\textbf{Missing}} & 
\multicolumn{1}{c}{\textbf{Tensor}}  & 
\multicolumn{1}{c}{\textbf{Missing}} & &
\multicolumn{2}{c}{\textbf{BAMITA Correlated}} & &
\multicolumn{1}{c}{\textbf{BAMITA Indep}} & 
\multicolumn{1}{c}{\textbf{missForest}} & &
\multicolumn{1}{c}{\textbf{EM}} & &
\multicolumn{1}{c}{\textbf{Corr}}& 
\multicolumn{1}{c}{\textbf{Inde}}& 
\multicolumn{1}{c}{\textbf{Overall}}\\ 
\cline{5-7}  
\multicolumn{1}{c}{\textbf{pattern}} & 
\multicolumn{1}{c}{\textbf{dimension}} & 
\multicolumn{1}{c}{\textbf{proportion}} & &
\multicolumn{1}{c}{\textbf{MSE}} & 
\multicolumn{1}{c}{\textbf{MSE low-rank}}& &
\multicolumn{1}{c}{\textbf{MSE}}&
\textbf{MSE} & &
\multicolumn{1}{c}{\textbf{Algorithm}}&&
\multicolumn{1}{c}{\textbf{prop}}&
\multicolumn{1}{c}{\textbf{prop}}&
\multicolumn{1}{c}{\textbf{prop}}\\

 & & & &  
\textbf{Coverage(\%)} & 
\textbf{Coverage(\%)} & & 
\textbf{Coverage(\%)} & 
\textbf{} & & 
\textbf{MSE} & & \textbf{(\%)} & \textbf{(\%)}&  \textbf{(\%)}\\ \hline

Entry Missing & (10$\times$ 10$\times$ 10) 
&   20\% & & {0.009} & 0.001 && 0.067 & 0.617 &  & 0.076  & & 100 &  99&  99\\
&   &     & & 97.4 & 98.2 && 94.4 &  &  & 0.076  & & 100 &  99&  99\\

& & 50\% & & {0.033} & 0.007 && 0.064 & 0.914&  & 0.110  & & 100 &  92&  92\\
&   &     & & 95.8 & 96.3 && 94.8 &  &  & 0.076  & & 100 &  99&  99\\

& & 70\% & & {0.102} & 0.044 && 0.113 & - &  & 0.415  & & 99 &  85&  84\\
&   &     & & 93.6 & 93.7 && 94.6 &  &  & 0.076  & & 100 &  99&  99\\

   \\
            & (20$\times$ 20$\times$ 20) 
&   20\% & & {0.007} & 0.001 && 0.105 & 0.371 &  & 0.105  & & 100 &  100&  100\\
&   &     & & 97.2 & 98.3 && 94.6 &  &  & 0.076  & & 100 &  99&  99\\

& & 50\% & & {0.033} & 0.002 && 0.107 & 0.561 &  & 0.107  & & 100 &  100&  100\\
&  &      & & 95.9 & 97.3 && 94.8 &  &  & 0.076  & & 100 &  99&  99\\

& & 70\% & & \bfseries{0.083} & 0.009 && 0.117 & 0.886 &  & 0.151  & & 100 &  82&  82\\
&  &      & & 94.4 & 95.2 && 94.7 &  &  & 0.076  & & 100 &  99&  99\\

   \\
& (65$\times$ 168$\times$ 6) 
&   20\% & &  0.050 & 0.001 && 0.273 & 0.364 &  & 0.287  & & 95 & 98&  93\\
&  &      & & 95.8 & 97.3 && 94.3 &  &  & 0.076  & & 100 &  99&  99\\

& & 50\% & &  0.138 & 0.005 && 0.275 & 0.397 &  & 0.277  & & 90 & 93&  84\\
&  &      & & 95.2 & 96.5 && 94.3 &  &  & 0.076  & & 100 &  99&  99\\

& & 70\% & &  0.232 & 0.013 && 0.293 & 0.494 &  & 0.300  & & 77 & 94&  72\\
&  &      & & 94.6 & 95.9 && 94.3 &  &  & 0.076  & & 100 &  99&  99\\

   \\
  
Fiber Missing & (10$\times$ 10$\times$ 10) 
&   20\% & &  0.025 & 0.002 && 0.065 & 0.655 &  & 0.098 & & 98 & 97&  95\\
&  &      & & 94.5 & 98.4 && 94.7 &  &  & 0.076  & & 100 &  99&  99\\

& & 50\% & &  0.085 & 0.017 && 0.083 & 0.988 &  & 0.339 & & 75 & 64&  54\\
&  &      & & 92.7 & 95.0 && 94.8 &  &  & 0.076  & & 100 &  99&  99\\

& & 70\% & & -     & -    && -     & -   &  & -     & & 9 & 16&  4\\
&  &      & & - & -  && - &   &  &    & & & & \\

   \\
            & (20$\times$ 20$\times$ 20) 
&   20\% & &  0.027 & 0.001 && 0.108 & 0.387 &  & 0.108 & & 100 & 100&  100\\
&  &      & & 93.2 & 98.3 && 94.8 &  &  & 0.076  & & 100 &  99&  99\\

& & 50\% & &  0.073 & 0.003 && 0.112 & 0.595 &  & 0.145 & & 96 & 88&  85\\
&  &      & & 91.4 & 96.9 && 94.8 &  &  & 0.076  & & 100 &  99&  99\\

& & 70\% & &  0.130 & 0.020 && 0.145 & 0.967 &  & 0.507 & & 80 & 36&  32\\
&  &      & & 90.2 & 92.4 && 94.4 &  &  & 0.076  & & 100 &  99&  99\\

   \\
            & (65$\times$ 168$\times$ 6) 
&   20\% & &  0.109 & 0.002 && 0.286 & 0.421&  & 0.283 & & 82 & 99&  81\\
&  &      & & 92.3 & 98.1 && 94.7 &  &  & 0.076  & & 100 &  99&  99\\

& & 50\% & &  0.316 & 0.082 && 0.309 & 0.515&  & 0.437 & & 82 & 87&  71\\
&  &      & & 91.9 & 96.0 && 94.4 &  &  & 0.076  & & 100 &  99&  99\\

& & 70\% & & 0.798 & 0.538 &&  0.555 & 0.904&  & 0.761 & & 59 & 49&  33\\
&  &      & & 90.0 & 90.0 && 93.1 &  &  & 0.076  & & 100 &  99&  99\\

   \\\hline
\end{tabular}}
\end{table}

\end{center}

\subsection{Simulation study 3: Imputation for function of a fiber }
In the third simulation study, we examine whether an arbitrary function of the entire fiber can be reasonably captured when the data is generated with a correlation structure. This is motivated by our data applications where, instead of focusing on the imputed data entries, we are interested in the alpha diversity calculated as a function of the entire mode-2 fiber for each subject and each time point (see section \ref{sec:5} for more detail). We argue that although the MSE of each imputed element is mostly adopted as the metric for evaluating the imputation performance, the ``structure" of the imputed data slice (i.e., fiber) is also of interest for downstream analyses of the tensor. 

The data generating mechanism is similar to that of the second simulation study except that the covariance matrix $\Sigma_1$ and $\Sigma_3$ are now identity matrices with dimensions $I_2$ and $I_3$ respectively, and $\Sigma_2$ is a matrix with diagonal elements being 1 and other non-diagonal elements being 0.15. Under the data-generating mechanism, we do not expect the Bayesian correlated algorithm to outperform the independent algorithm in terms of the point-wise imputation error since the other modes (modes 1 and 3) are uncorrelated. However, with the additional model of the covariance structure for mode 2, our correlated algorithm may better capture the variation of the function of the imputed mode 2 fiber. To evaluate this, for each imputed fiber $\hat{\pmb{\mathscr{X}}}[i,\cdot,k]$, we calculate the linear predictor $\beta_b\hat{\pmb{\mathscr{X}}}[i,\cdot,k]$ with randomly generated coefficient $\beta_b$, $b=1,...,100$. Then we evaluate the mean MSE and coverage (only for fiber-wise missing) for the linear predictors over the 100 randomly generated coefficients $\beta_b$. The results are displayed as``MSE(Fiber)" in Table \ref{Table3}.

 The results are shown in Table~\ref{Table3}, from which we can see that, although the Bayesian independent algorithm has a relatively good coverage rate for each imputed element, the mean coverage for the random linear combination of each imputed fiber is not ideal, especially in the high-dimension cases. The Bayesian correlated algorithm, on the other hand, has substantially better coverage performance in terms of the imputed fiber.

\begin{center}
\begin{table}[]
\caption{Simulation results for study 3: imputation for a function of a fiber. Results are presented as median values over all the missing elements.\label{Table3}}
\resizebox{\textwidth}{!}{
\begin{tabular}{cccccccccccccc}
\hline
\multicolumn{1}{c}{\textbf{Missing}} & 
\multicolumn{1}{c}{\textbf{Tensor}}  & 
\multicolumn{1}{c}{\textbf{Missing}} & &
\multicolumn{2}{c}{\textbf{BAMITA}} & &
\multicolumn{2}{c}{\textbf{BAMITA}} & & 
\multicolumn{1}{c}{\textbf{EM Algorithm}} &  
\multicolumn{1}{c}{\textbf{Converged}}&  
\multicolumn{1}{c}{\textbf{Converged}}&  
\multicolumn{1}{c}{\textbf{Converged}}\\ 

\multicolumn{1}{c}{\textbf{pattern}} & 
\multicolumn{1}{c}{\textbf{dimension}} & 
\multicolumn{1}{c}{\textbf{proportion}} & &
\multicolumn{2}{c}{\textbf{Correlated}} &&
\multicolumn{2}{c}{\textbf{Independent}}&&
\multicolumn{1}{c}{\textbf{True Rank}}&
\multicolumn{1}{c}{\textbf{Proportion}}&
\multicolumn{1}{c}{\textbf{Proportion}}&
\multicolumn{1}{c}{\textbf{Proportion}}\\

\cline{5-8} \cline{9-10} & & & &  
\textbf{MSE(Imputation)} & 
\textbf{MSE(Fiber)} & & 
\textbf{MSE(Imputation)} & 
\textbf{MSE(Fiber)} & & 
\textbf{MSE(Imputation)}  &
\textbf{Correlated}  &
\textbf{Independent} &
\textbf{Overall} \\

 & & & &  
\textbf{Coverage(\%)} & 
\textbf{Coverage(\%)} & & 
\textbf{Coverage(\%)} & 
\textbf{Coverage(\%)} & & 
\textbf{-} &  
\textbf{-}&  
\textbf{-}&  
\textbf{-} \\ \hline

Entry Missing & (10$\times$ 10$\times$ 10) 
&   20\% & & 0.409 & - && \bfseries0.314 & - &  & 0.322 & 100& 100& 100\\
& &   & & 88.0 & \bfseries78.0 && \bfseries95.0 & 53.0 &  & - \\

& &  50\% & & 0.396 & - && \bfseries0.327 & - &  & 0.423 & 100& 97& 97\\
& &   & & 91.3 & \bfseries90.0 && \bfseries95.0 & 72.0 &  & - \\

& &  70\% & & 0.596 & - && \bfseries0.466 & - &  & 0.776 & 98& 99& 97\\
& &   & & 90.7 & \bfseries89 && \bfseries94.9 & 83.5 &  & - \\
   \\
            & (20$\times$ 20$\times$ 20) 
&   20\% & & 0.266 & - && \bfseries0.260 & - &  & 0.265 & 100& 100& 100\\
& &   & & 92.2 & \bfseries91.2 && \bfseries94.8 & 36.2 &  & - \\

& &  50\% & & 0.284 & - && \bfseries0.274 & - &  &  0.282 & 100& 100& 100\\
& &   & & 92.0 & \bfseries91.2 && \bfseries94.9 & 53.2 &  & - \\

& &  70\% & & 0.278 & - && \bfseries0.272 & - &  &  0.326 & 100& 86& 86\\
& &   & & 93.0 & \bfseries91.0 && \bfseries94.9 & 62.3 &  & - \\
   \\
            & (65$\times$ 168$\times$ 6) 
&   20\% & &  0.287 & - && \bfseries0.260 & - &  & 0.270 & 100& 99& 99\\
& &   & & 91.9 & \bfseries91.8 && \bfseries94.8 & 13.8 &  & -\\

& &  50\% & & \bfseries0.268 & - && 0.271 & - &  & 0.299 & 100& 98& 98\\
& &    & & 92.9 & \bfseries92.6 && \bfseries94.8 & 22.3 &  & - \\

& &  70\% & & \bfseries0.250 & - &&  0.278 & - &  & 0.305 & 100& 96& 96\\
& &   & & 93.6 & \bfseries92.1 && \bfseries94.7 & 26.0 &  & - \\

Fiber Missing & (10$\times$ 10$\times$ 10) 
&   20\% & & 0.404 & 0.767 && \bfseries0.311 &\bfseries 0.617 &  & 0.332 & 99& 100& 99\\
& &   & & 90.0 & \bfseries90.2 && \bfseries94.8 & 83.0 &  & - \\

& &  50\% & & 0.875 & 1.122 && \bfseries0.514 & \bfseries0.915 &  & 0.612 & 78& 92& 73\\
& &   & & 89.5 & \bfseries87.1 && \bfseries94.6 & 82.8 &  & - \\

& &  70\% & & \bfseries2.609 & 5.153 && 4.041 & \bfseries3.510 &  & 0.964 & 9& 39& 7\\
& &   & & 86.6 & 83.8 && \bfseries93.2 & \bfseries84.2 &  & - \\
   \\
            & (20$\times$ 20$\times$ 20) 
&   20\% & & 0.285 & 0.668 && \bfseries0.266 & \bfseries0.616 &  & 0.273 & 100& 100& 100\\
& &   & & 93.0 & \bfseries90.3 && \bfseries94.9 & 67.9 &  & - \\

& &  50\% & & 0.320 & 0.766 && \bfseries0.285 & \bfseries0.701 &  & 0.367  & 100 & 91 & 91\\
& &   & & 92.3 & \bfseries87.4 && \bfseries94.8 & 68.4 &  & -\\

& &  70\% & & 0.430 & 0.837 && \bfseries0.369 & \bfseries0.769 &  & 0.645 & 93& 61& 59\\
& &   & & 91.8 & \bfseries86.1 && \bfseries94.6 & 69.6 &  & - \\
   \\
            & (65$\times$ 168$\times$ 6) 
&   20\% & & 0.261 &\bfseries 0.940 && \bfseries0.259 & 0.947 &  & 0.288 & 100& 99& 99\\
& &   & & 94.6 & \bfseries80.0 && \bfseries94.8 & 29.4 &  & - \\

& &  50\% & & 0.465 & 0.990 && \bfseries0.337 &\bfseries 0.958 &  & 0.431 & 89& 85& 77\\
& &    & & 92.9 & \bfseries76.3 &&\bfseries 94.6 & 31.0 &  & - \\

& &  70\% & & 0.869 & 1.005 && \bfseries0.540 & \bfseries0.979 &  & 0.686 & 32& 43& 15\\
& &   & & 90.7 &\bfseries 71.8 && \bfseries93.2  & 32.0 &  & - \\
   \\\hline
\end{tabular}
}
\end{table}
\end{center}

\section{Longitudinal microbiome applications}
\label{sec:5}

\subsection{Infant gut microbiome application}
 \label{infant.app}

The gut houses a rich array of microbial organisms, presenting a diverse landscape. This dynamic ecosystem holds potential as significant indicators for both digestive and broader health conditions. We consider a longitudinal study of the gut microbiome on 52 infants in the neonatal intensive care unit (NICU), where stool samples were collected over the first 3 months of life \citep{cong2017influence}.   Using 16s rRNA sequencing technology, we obtained microbiome data, which we aggregated to the genus level, yielding 152 distinct genera.  Employing standard preprocessing techniques, we addressed zero values by introducing pseudo counts, transformed the data into compositional profiles, and applied the centered log-ratio (clr) transformation. Data were aggregated every 5 consecutive days, yielding 30 time intervals. Consequently, we obtained a tensor data array with dimensions of $52\times 152 \times 30$. However, due to the unavailability of samples from every infant at every time point, the tensor array exhibits a fiber-wise missing structure, with approximately 71\% of the samples being absent. Our objective is to employ BAMITA to address missing values and assess the dynamic diversity in the microbiome over this population. Before applying our algorithm to this dataset, we first check the suitability of the normality assumption and the separable covariance assumption. Results are presented in Appendix~\ref{appA}, and suggest that the two assumptions are reasonable for these data. \color{black}
The microbiome diversity is often measured using alpha diversity which refers to diversity on a local scale. \citep{thukral2017review}.  Here, we compute diversity using the Shannon-Wiener index \citet{shannon1948mathematical}. For a given fiber of the data X$_{i,,k}$, the Shannon alpha diversity is calculated as $-\sum_{j=1}^{152} p_j\log(p_j)$ where $p_j=\frac{exp(\textnormal{ClrX}_{i,j,k})}{\sum_{j=1}^{152}exp(\textnormal{ClrX}_{i,j,k})}$ is the proportion of element $j$ in the sample. 

We impute the missing fibers of the microbiome tensor array with the two proposed Bayesian algorithms and the EM algorithm approach. For the Bayesian multiple imputation algorithm with separable covariance structure error, we assume independence in the first dimension (the 52 infants) but account for correlation across time and genera. The performance of the different algorithms is evaluated through cross-validation. 
 For each of $200$ simulation experiments, we randomly hold an additional $25\%$ of the observed fibers out and calculate the mean squared error of the imputed values and the true observed value. 
Shannon diversity is also computed for the missing fibers at each MCMC iteration, and they are compared to the observed diversity measures for the validation set. For the Bayesian imputation with covariance structure (BAMITA Correlated), we also evaluate the MSE for the estimated low-rank structure, i.e., $\llbracket {(\mathbf{U}^{1})},..., (\mathbf{U}^{N}) \rrbracket $, which represent the imputation without adjustment of covariance structure.

The results are summarized in 
Table~\ref{Table_ClrX2}.  The low-rank Bayesian imputation approach with correlation structure performs substantially better than other approaches with respect to MSE for both imputed values in the fiber and Shannon diversity. This suggests that the data have a strong low-rank structure, with substantial correlation in the residual covariance. Moreover, while coverage rates are appropriate for the fiber-wise entries under both models, coverage for Shannon diversity is much higher for the correlated model.  This illustrates the notion that accurately inferring uncertainty in the marginal distribution for the entries of an array does not imply uncertainty will be accurately inferred for multivariate functions that are used in downstream analysis. The MSE for the low-rank structure in the correlated algorithm is similar to the MSE of the independent algorithm, indicating that adjusting the covariance structure helps the imputation performance. 

\begin{center}
\begin{table}[!h]
\caption{Fiber-wise imputation results under cross-validation for the neonatal ClrX application.  'Imputation MSE' gives relative MSE for held-out values in the tensor, 'low-rank MSE' gives relative MSE when imputing via the low-rank term only in the correlated model, and 'Shannon MSE' gives MSE for imputed Shannon entropy for held-out fibers. Coverage rates for 95\% credible intervals are shown in parentheses.      \label{Table_ClrX2}}
\resizebox{\textwidth}{!}{
\begin{tabular}{ccccccccccc}
\hline

\multicolumn{1}{c}{\textbf{Number of}} & &
\multicolumn{3}{c}{\textbf{BAMITA}} &  & &
\multicolumn{2}{c}{\textbf{BAMITA}}  & &
\multicolumn{1}{c}{\textbf{EM Algorithm}} \\

\multicolumn{1}{c}{\textbf{components}} & &
\multicolumn{3}{c}{\textbf{Correlated}}  & & &
\multicolumn{2}{c}{\textbf{Independent}} & &
\multicolumn{1}{c}{\textbf{}}\\

\cline{3-5} \cline{7-10}  & &  
\textbf{Imputation MSE} & 
\textbf{Low-rank MSE} & 
\textbf{Shannon MSE} &
& &
\textbf{Imputation MSE} & 
\textbf{Shannon MSE} &
 &
\textbf{Imputation MSE}   \\ 
& &
\textbf{(Coverage)} & 
\textbf{} & 
\textbf{(Coverage)} & 
  & &
\textbf{(Coverage)} & 
\textbf{(Coverage)}  & & 
\textbf{}

\\\hline










  
  1 & & 0.340 & 0.583 & 0.744  &&& 0.560 & 1.834 &   & 0.569 \\
&& (95.4) &   & (84.3)  &&& (94.1) & (55.3) &   &   \\

  2 & & 0.345 & 0.570 & 0.698 & && 0.533 & 1.471 &  & 0.557 \\
&& (95.3) &   & (84.4)& && (94.3) & (55.0)&   &   \\

  3 & & 0.353 & 0.561 & 0.665 & && 0.518 & 1.035 &  & 0.554 \\
&& (95.1) &   & (84.2)& && (94.5) & (63.0)&   &   \\

  4 & & 0.365 & 0.547 & 0.635 & && 0.505 & 0.924 &   & 0.543 \\
&& (95.0) &   & (81.9)& && (94.5) & (63.3)&   &   \\

  5 & & 0.384 & 0.552 & 0.611 & && 0.505 & 0.903  &  & 0.555 \\
&& (94.8) &   & (82.0)&&& (94.3) & (61.7)&   &   \\

 6 & & 0.391 & 0.547 & 0.608 & && 0.497 & 0.874 &   & 0.554 \\
&& (94.6) &   & (81.8)& && (94.2) & (60.0)&   &   \\

  7 & & 0.423 & 0.576 & 0.608 & && 0.496 & 0.827 &  & 0.555 \\
&& (94.3) &   & (80.9) &&& (94.1) & (59.7)&   &   \\

  8 & & 0.439 & 0.578 & 0.605 & && 0.499 & 0.794 &   & 0.554 \\
&& (94.2) &   & (80.8)&&& (94.1) & (59.1)&   &   








   \\\hline
\end{tabular}
}
\end{table}
\end{center}

We apply the rank-1 correlated model to the full data, with no validation set, to infer trends in microbiome diversity over time.  We consider three approaches to generate uncertainty bounds for the mean diversity at each time point.  For approach 1., we impute the missing data at their posterior mean, treat the values as fixed, and create 95\% confidence intervals using the classical frequentist approach via a t-distribution.  For approach 2., we use only the observed data at each timepoint, and create a 95\% interval via a t-distribution.  Note the t-interval for a mean is equivalent to a Bayesian credible interval with a uniform prior on the mean and log-uniform prior on the variance of the values.  Thus, for approach 3. we use the posterior samples to propagate uncertainty from the imputation step and generate credible intervals for the full model.  That is, let $\text{mean}(\alpha_{k,t})$ and $\text{sd}{(\alpha_{k,t})}$ be the sample mean and standard deviation for diversity at timepoint $k$ and MCMC iteration $t$, including observed values and imputed values simulated from the posterior. Then, we simulate a value for the  population mean $\bar{\alpha}_{k,t}$ via
\[\bar{\alpha}_{k,t}=\text{mean}(\alpha_{k,t})+\text{sd}{(\alpha_{k,t})} T_{51}\]
where $T_{51}$ is a t-distributed random variable with 51 degrees of freedom. Intervals obtained via the quantiles of the $\bar{\alpha}_{k,t}$ will then properly account for both variability in the imputed values  and sampling variability.    

\begin{figure}[!h]
\caption{Estimates for Shannon diversity over time for the neonatal infant ClrX data. The left panel gives the mean under imputation and credible intervals generated using either point or multiple imputation, the right panel gives the mean and credible interval for observed data only.}
\label{fig_diversity2}
\includegraphics[width=\textwidth]{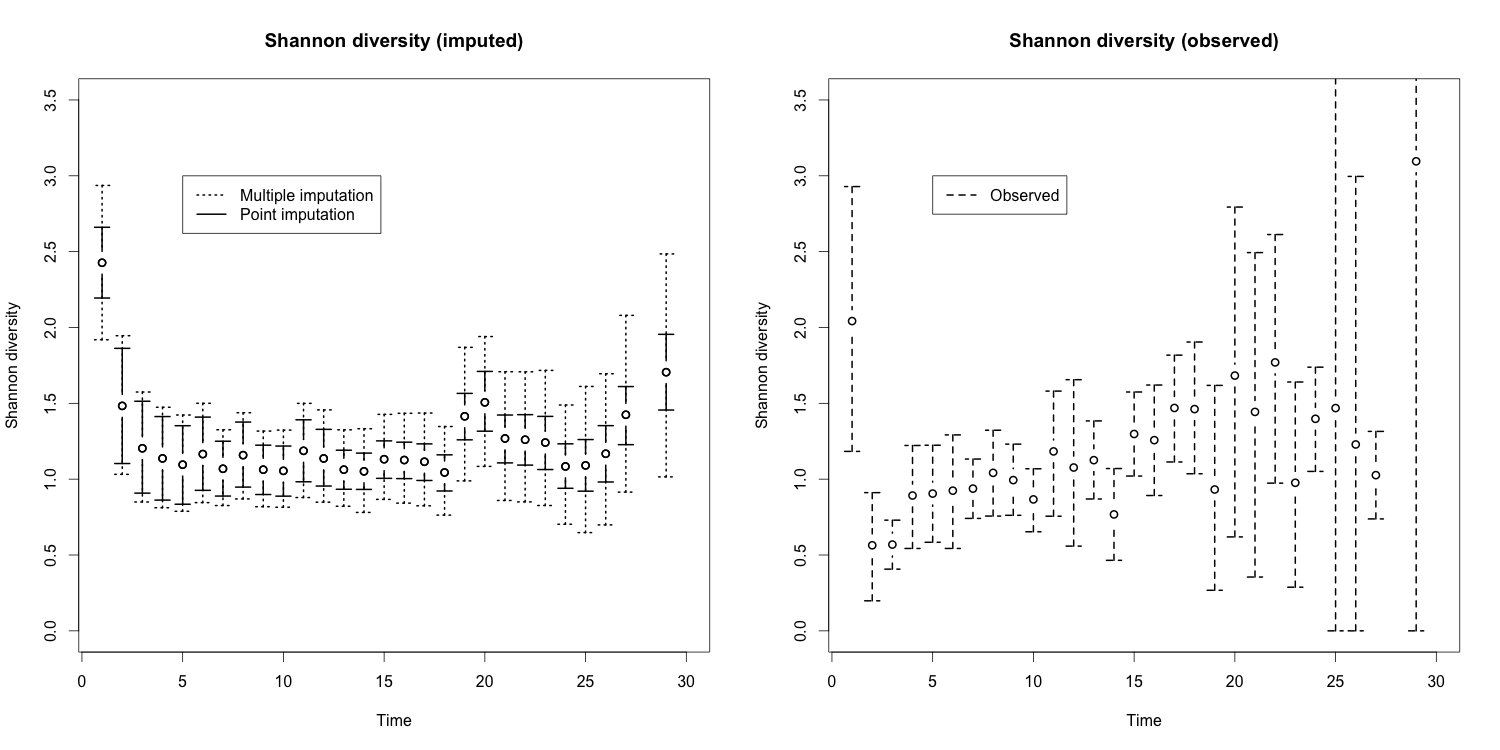}
\end{figure}

The resulting trends for Shannon diversity, with 95\% confidence/credible bounds, are shown in Figure~\ref{fig_diversity2}.  The confidence bounds generated using the point-imputed data are substantially more narrow than the bounds generated using multiple imputation in some instances, illustrating the danger of underestimating uncertainty if imputed data are treated as fixed and known. In contrast, the bounds generated using only the observed data are sporadic and very wide in some cased, illustrating the disadvantages of completely ignoring missing data. The multiple imputation approach is a principled compromise between these two extremes, and show a pattern in which diversity decreases over the first few days of life in the NICU and then stays relatively constant. 

\subsection{Mouse oral microbiome application}
Here we consider a longitudinal study of the oral microbiome, where we examined saliva samples collected from mice induced with carcinogens. Throughout the study, we tracked 65 mice across 6 regular sampling intervals. Data were processed as in Section \ref{infant.app}, yielding a tensor with dimensions of $65\times 168 \times 6$. Due to the unavailability of saliva samples for all mice at every time point, this tensor also has fiber-wise missingness with approximately 25\% of the samples being absent. 

We evaluate performance for fiber-wise missing imputation as in Section~\ref{infant.app} through cross-validation, and for illustration purposes we also consider performance of entry-wise imputation in which we randomly hold an additional $25\%$ of the observed entries in the tensor out. 

The results are summarized in Table~\ref{Table_ClrX}.  For both scenarios, the Rank-1 Bayesian imputation approach with correlation structure performs better than or similarly to alternatives. The fiber-wise MSE is similar between the correlated and independent methods. However, the MSE for Shannon diversity and  entry-wise imputation is improved by accounting for correlation, are improved by accounting for residual correlation.  Moreover, the model with residual correlation structure has appropriate coverage rates for the Shannon diversity, whereas the model with independence has under-coverage.     However, the Bayesian model with residual correlation structure has appropriate coverage rates for the two diversity measures, whereas the model with independence has under-coverage.  This again illustrates the importance of accurately capturing the multivariate distribution to reflect uncertainty in downstream analyses.  Note, however, that the correlated model performs substantially worse for higher ranks with fiber-wise missingness, and this is likely due to overfitting with the large number of parameters. 

\begin{figure}[!h]
\caption{Estimates for mean Shannon diversity  over time for the mouse application, with credible intervals generated using either point imputation, multiple imputation, or the observed data only.}
\label{fig_diversity}
\includegraphics[width=\textwidth]{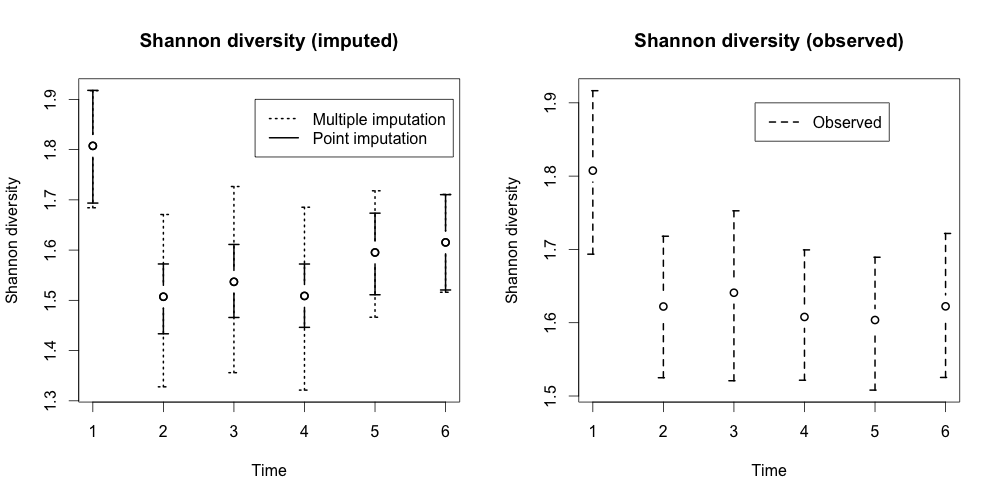}
\end{figure}

We apply the rank-1 correlated model to the full data to infer trends in microbiome diversity, using the same approaches described in Section 5.1 of the manuscript.   The resulting trends, with 95\% confidence/credible bounds, are shown in Figure~\ref{fig_diversity}.  The confidence bounds generated using the point-imputed data are again much more narrow than the bounds generated using multiple imputation, illustrating the danger of underestimating uncertainty if imputed data are treated as fixed and known.  Nevertheless, the bounds generated using multiple imputation (and using only the observed data) still show significant changes over time. Namely,  Shannon diversity decreases sharply after exposure, and there is some evidence for a rebound at subsequent timepoints but it is not conclusive.

\begin{center}
\begin{table}[!h]
\caption{Imputation performance under cross-validation for the mouse ClrX application. 'Imputation MSE' gives relative MSE for held-out values in the tensor, 'low-rank MSE' gives relative MSE when imputing via the low-rank term only in the correlated model, and 'Shannon MSE' gives MSE for imputed Shannon entropy for held-out fibers. Coverage rates for 95\% credible intervals are shown in parentheses.  \label{Table_ClrX}}
\resizebox{\textwidth}{!}{
\begin{tabular}{cccccccccccc}
\hline
\multicolumn{1}{c}{\textbf{Missing}} & 
\multicolumn{1}{c}{\textbf{Number of}} & &
\multicolumn{3}{c}{\textbf{BAMITA}} & &  &
\multicolumn{2}{c}{\textbf{BAMITA}} &  &
\multicolumn{1}{c}{\textbf{EM Algorithm}} \\ 

\multicolumn{1}{c}{\textbf{pattern}} & 
\multicolumn{1}{c}{\textbf{components}} & &
\multicolumn{3}{c}{\textbf{Correlated}} &  & &
\multicolumn{2}{c}{\textbf{Independent}}&  &
\multicolumn{1}{c}{\textbf{}}\\

\cline{3-7} \cline{9-11}   & & &  
\textbf{Imputation MSE} & 
\textbf{Low-rank MSE} & 
\textbf{Shannon MSE}  & & &
\textbf{Imputation MSE} & 
\textbf{Shannon MSE}  & & 
\textbf{Imputation MSE}   \\ 
& & &
\textbf{(Coverage)} & 
\textbf{} & 
\textbf{(Coverage)}  & & &
\textbf{(Coverage)} & 
\textbf{(Coverage)}  & & 
\textbf{}

\\\hline

Random Missing 
&  1  & & 0.261 & 0.315 & -  &&& 0.315 & -  &  & 0.316  \\
&&& (96.1) &   & -  &&& (94.9) & - &   &    \\

&  2  & & 0.259 & 0.304 & -  &&& 0.295 & -  &  & 0.306  \\
&&& (96.0) &   & -  &&& (94.9) & - &   &   \\

&  3  & & 0.259 & 0.294 & -  &&& 0.285 & - &   & 0.299 \\
&&& (96.0) &   & -  &&& (94.9) & - &   &   \\

&  4  & & 0.268 & 0.297 & -  &&& 0.280 & - &   & 0.296 \\
&&& (96.0) &   & -  &&& (94.9) & - &   &   \\






  
Fiber Missing 
&  1 & & 0.338 & 0.341 & 0.117  &&& 0.339 & 0.169  & & 0.338 \\
&&& (96.6) &   & (97.5) &&& (94.8) & (87.0) &  &   \\

&  2 & & 0.969 & 0.971 & 0.105  &&& 0.394 & 0.139  &  & 0.361 \\
&&& (96.3) &   & (96.0)& && (94.8) & (82.6)&   &   \\

&  3 & & 3.291 & 3.293 & 0.151 1 &&& 0.632 & 0.129 &   & 0.377 \\
&&& (94.4) &   & (90.8)& && (94.9) & (81.6)&   &   \\

&  4 & & 7.306 & 7.306 & 0.234  &&& 0.843 & 0.127 &  & 0.450 \\
&&& (94.1) &   & (82.6)& && (94.0) & (78.5)&   &   \\

   \\\hline
\end{tabular}
}
\end{table}
\end{center}

\section{Discussion}

Our results demonstrate the advantages of accounting for residual covariance and uncertainty when imputing missing values in tensor data.  While the motivating application for the development of BAMITA was longitudinal microbiome data, the model is generally applicable to a wide variety of application scenarios.  Aspects of the model and sampling algorithm may be modified or extended, e.g., to capture spatiotemporal structure in relevant modes \citep{yokota2016smooth, guan2023smooth} rather than a general covariance.  Moreover, the assumption of normality for the residual error may be relaxed. For example, a tensor modeling approach is often effective for multi-condition RNA-Seq gene expression data \citep{hore2016tensor}, which may have a Poisson or negative binomial distribution.

\
In our data applications we selected the rank using cross-validation  (Algorithm \ref{alg2}). Thus, subsequent analyses with the selected rank for the same data are prone to post-selection inference.  However, as just one parameter (the rank) is empirically estimated, over fitting is not a major concern. For example, in our simulation, coverage rates are appropriate when the rank is selected through cross-validation. Nevertheless, a possible extension is to infer the number of components $R$ (i.e., rank) as a parameter within the Bayesian model to account for its uncertainty, using reversible jump MCMC 
\citep{fruhwirth2024sparse} or other approaches.  However, this would increase the computational complexity of the approach.

To make our algorithm applicable in various data application scenarios, we adopted the flat prior for the underlying parameters $\boldsymbol{U}$ which is independent of the data scale. However, under the specific data application scenarios, informative Gaussian priors can also be considered in our algorithm as a conjugate prior. One future topic is to extend our algorithm with the informative Gaussian priors on the underlying structure elements $\boldsymbol{U}^{(1)}, \boldsymbol{U}^{(2)}, ..., \boldsymbol{U}^{(N)}$. 


In our real data applications, we use  cross-validation to decide whether to adopt the Bayesian multiple imputation algorithm with the separable covariance structure or the independent structure, using mean squared error (MSE) of the imputed elements. However, alternative Bayesian model selection criteria, such as the deviance information criterion (DIC) \citep{spiegelhalter2002bayesian}, could also be used to select the most appropriate model. 

\color{black}
Computing time is often a bottleneck to fully Bayesian inference for high-dimensional data. We have carefully specified our models to facilitate efficient Gibbs sampling in high-dimensions.  However, the model with independence error allows for a much more efficient algorithm; our largest dataset, described in Section~\ref{infant.app}, took several hours to run under the correlated model and under 10 minutes with independent error. Thus, this is a trade-off to modeling the residual covariance in higher dimensions.

\section*{Acknowledgments}
The authors thank the anonymous reviewers for their valuable suggestions. This work was supported in part by NIH grants R01-HG010731 and R01-GM130622.


\appendix
\section{Additional plots for microbiome applications}
\label{appA}

Here we present some plots for the data used in both microbiome applications presented.  In particular,  we check the appropriateness of the normality assumption and a separable covariance structure for both  data applications. 

\begin{figure}[!h]
    \centering
    \caption{Histograms for the standardized data from the two case studies}
     \includegraphics[width=\textwidth]{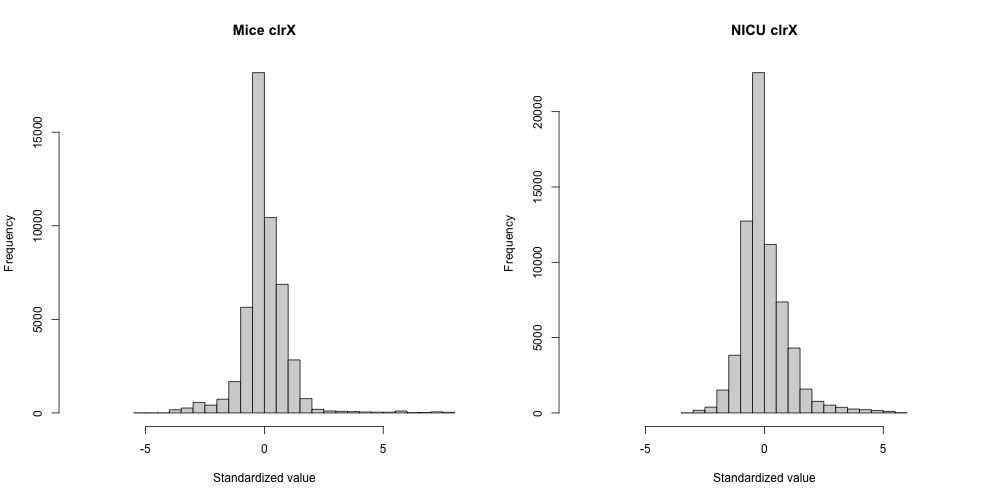}
    \label{fig:hist}
\end{figure}

We first present the histogram for the standardized observed data values for both the NICE data and the mice data. The histograms are present in Figure \ref{fig:hist}. From the figure, we can see that the histograms are generally centered around zero and exhibit a bell-shaped distribution without significant skewness. This suggests that, while the data may not be perfectly normal, the model with normally distributed errors can be a reasonable approximation for both datasets.  This is further validated supported by the appropriate coverage rates for held-out data values shown for both applications. 

\begin{figure}[!h]
    \centering
    \caption{Heatmaps for the empirical taxa correlation matrix of the NICU example at different observation timepoints.}
     \includegraphics[width=\textwidth]{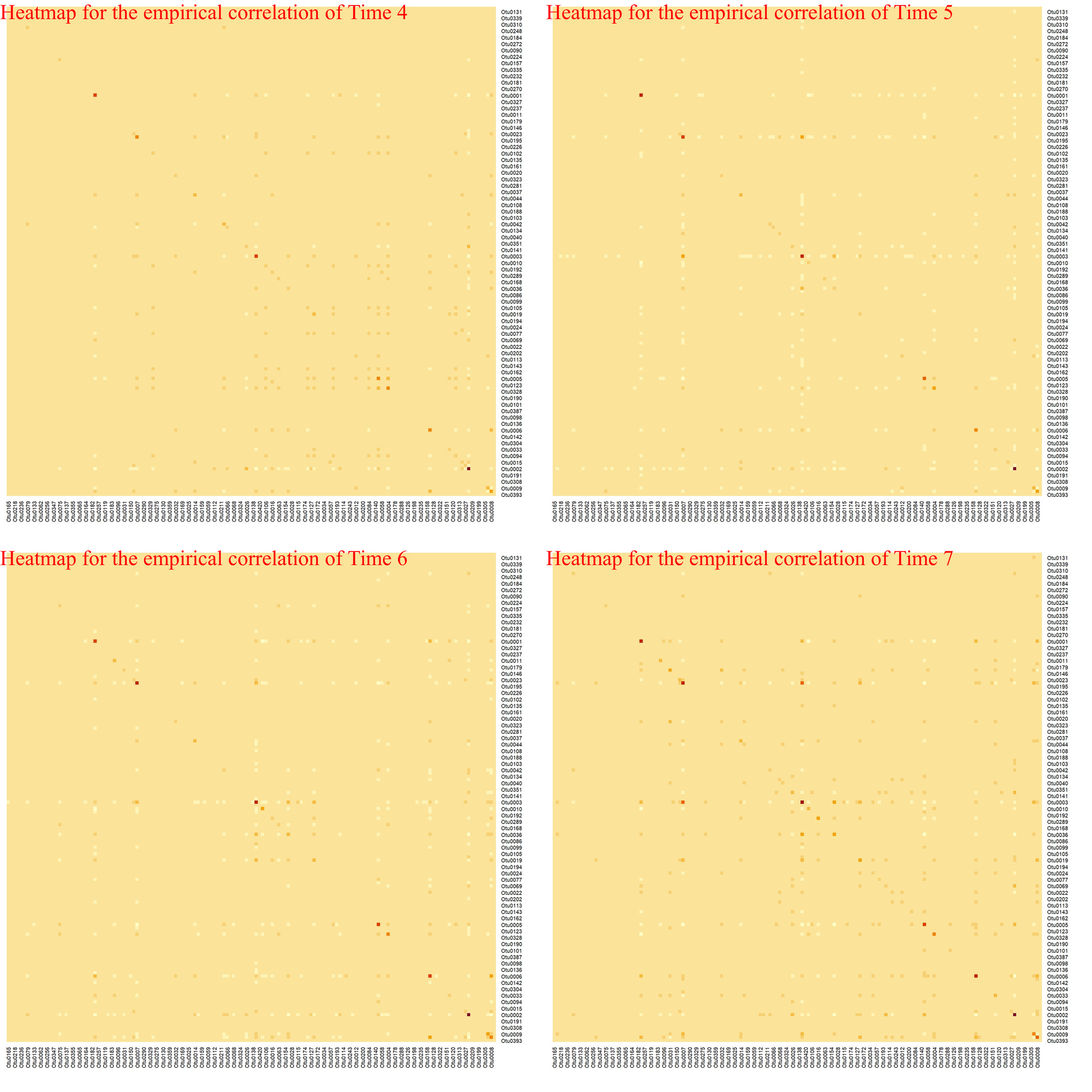}
    \label{fig:heatnicu}
\end{figure}

\begin{figure}[!h]
    \centering
    \caption{Heatmaps for the empirical taxa correlation matrix of the mice example at different observation timepoints.}
     \includegraphics[width=\textwidth]{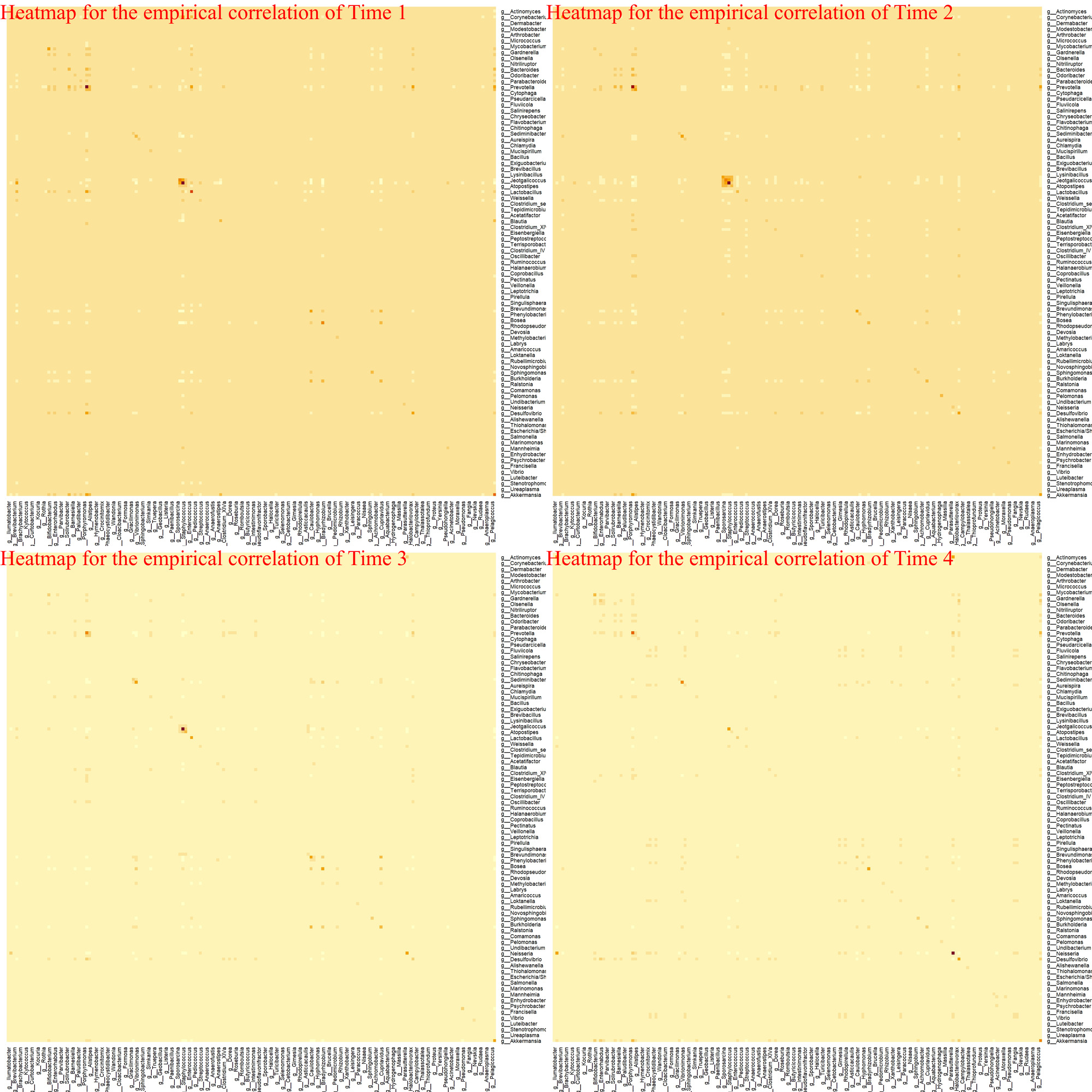}
    \label{fig:heatmice}
\end{figure}

In Figure \ref{fig:heatnicu} and \ref{fig:heatmice}, we present the heatmap for the  correlation matrix of the observed data across taxa at different time points. From the figures, we can see that the heatmaps are very similar at different time points. The consistency of these heatmaps supports the assumption that the relative correlation structure remains the same over time, as is assumed for a separable covariance.

\clearpage
\bibliographystyle{Chicago}
\bibliography{Bibliography}

\end{document}